\documentclass[aps,prb,reprint,showpacs,amsmath,amssymb]{revtex4-2}
\usepackage{graphicx} 
\usepackage{hyperref} 
\usepackage{bm}       
\usepackage{physics}
\usepackage[dvipsnames]{xcolor}
\definecolor{mypine}{RGB}{1, 121, 111}

\begin{document}


\title{Disorder-induced strong-field strong-localization in 2D systems}

\author{Yi Huang}
\affiliation{Condensed Matter Theory Center and Joint Quantum Institute, Department of Physics, University of Maryland, College Park, Maryland 20742, USA}

\author{Sankar Das Sarma}
\affiliation{Condensed Matter Theory Center and Joint Quantum Institute, Department of Physics, University of Maryland, College Park, Maryland 20742, USA}

\date{\today}


\begin{abstract}
A recent STM experiment in 2D bilayer graphene [Y.-C. Tsui, \textit{et al.}, \href{https://doi.org/10.1038/s41586-024-07212-7}{Nature \textbf{628}, 287 (2024)}], under a strong perpendicular magnetic field, has made a direct observation of the existence of three distinct filling-factor-dependent quantum phases in the lowest Landau level: the incompressible fractional quantum Hall liquid, a crystalline compressible hexagonal Wigner crystal with long-range order and rotational symmetry-breaking, and a random localized solid phase with no spatial order. We argue that the spatially random localized phase at low filling is the recently proposed disorder-dominated strongly localized amorphous ``Anderson solid'' phase [A. Babber, \textit{et al.}, \href{https://arxiv.org/abs/2601.03521}{arXiv:2601.03521}], which appears generically at a sample-dependent filling factor.
\end{abstract}

\maketitle


\section{Introduction}

In a breakthrough recent STM experiment \cite{Yazdani2024}, 2D-confined electrons under a strong magnetic field are directly imaged in Bernal bilayer graphene (BLG) with the spectacular direct manifestations of the fractional quantum Hall liquid (FQHL), the Wigner crystal (WC), and an amorphous solid (AS) phase at various fillings ($\nu$) of the lowest Landau level (LLL). Other than the considerable significance of direct imaging of electrons showing their spatial structure and organization in highly correlated FQHL and WC phases, there are several important features in the data elucidating deep physics.

First, the system generically displays a hexagonal WC with explicit breaking of the rotational symmetry over a broad range of fillings, and an incompressible FQH liquid appears only in a very narrow window around the odd-denominator fraction such as $\nu=1/3$ Laughlin state (e.g., $\nu \approx 0.334$). Even a slight deviation to $\nu \approx 0.311$ or $0.356$ rapidly restores the WC pattern, consistent with the idea that the $\nu=1/3$ phase is stabilized only within a narrow range close to commensurate odd-denominator fillings. The FQH state slightly lowers its energy compared with the background of a smooth WC phase (existing for all values of $\nu$) through small cusps at these precise odd-denominator $\nu$ (e.g., $1/3$ real space profile as can be seen clearly in Fig.~1 of Ref.~\onlinecite{Yazdani2024}). Although this is not theoretically unexpected \cite{Yoshioka1983, Khrapai2008, Yacoby2012}, the observation of a generic WC surrounding the $\nu=1/3$ state~\cite{Yazdani2024} provides direct verification of this theoretical expectation that FQHL states are stabilized only within a narrow window centered on odd-denominator fillings.
This experiment~\cite{Yazdani2024} is also the first direct observation of the generic existence of the WC phase in the lowest Landau level of 2D systems.

A new experimental finding in Ref.~\onlinecite{Yazdani2024}, however, is that, as $\nu$ continues decreasing, eventually at a small $\nu = 0.093 \approx 1/11$, the electrons cross over to a completely spatially random amorphous solid phase. 
In the WC, the structure factor $S(q)$ exhibits six well-developed Bragg peaks, reflecting the long-range hexagonal order of the crystal, whereas in the FQHL $S(q)$ is essentially a featureless blob, consistent with the absence of long-range spatial correlations in real space. By contrast, in the amorphous solid phase the real-space images show electrons localized at random, quenched positions, and the corresponding $S(q)$ displays a characteristic hybrid structure: a diffuse central blob accompanied by a surrounding ring. The presence of the ring in $S(q)$ unambiguously identifies the amorphous solid as a solid with a well-defined characteristic length scale, while the central blob signifies liquid-like, short-range spatial correlations. Together, these features establish the amorphous solid as a distinct phase: solid in the sense of electron localization, yet lacking the long-range translational and rotational order characteristic of a crystalline WC.
Another impressive recent zero-field density-tuned STM study has reported the observation of a  low-density zero-field electronic amorphous phase in highly disordered bilayer MoSe$_2$ samples (no moiré)~\cite{ge2025visualizingimpactquencheddisorder}, which has been interpreted as the manifestation of the strongly-localized random disorder-dominated zero-field AS phase in Ref.~\cite{babbar2026wignersolidandersonsolid}.

The transition from the $\nu=0.334$ FQHL to the $\nu=0.093$ AS through the WC phase for $0.083 < \nu < 0.334$ is a gradual crossover, and not a sudden sharp transition (we defer to Ref.~\onlinecite{Yazdani2024}, particularly its Fig.~2, for the details). In the current work, we focus on the amorphous solid phase at $\nu \approx 0.093$, and argue that this is Anderson (and amorphous) solid phase (AS), where disorder drives the electrons into a strongly localized random pattern exhibiting a glassy structure factor. This AS phase is a highly disordered strongly localized insulator, adiabatically connected to the Anderson localization fixed point \cite{Anderson1958}, where both $\sigma_{xx}$ and $\sigma_{xy}$ vanish in transport measurements at some nonuniversal sample dependent filling factor $\nu_c$.
Transport studies \cite{Shayegan1997,Fertig1997, Wilson1981,Willett1988,Ye2002,Goldman1990,Manoharan1996,Madathil:2024,Wang2025,Wang:2025b} have almost exclusively described the low-filling insulating phase as a pinned WC, but Ref.~\onlinecite{Yazdani2024} raises serious questions with respect to this pinned WC interpretation since the STM images indicate a generic existence of the WC for the whole filling factor range $1/3 > \nu > \nu_c$ ($\sim 1/11$). 
Since the disorder landscape of the sample is fixed, a natural question arises as to why WC pinning would occur only at the specific filling $\nu \approx 1/11$, and not, for example, at $\nu \approx 0.243$, where disorder is already evident, the electrons are strongly distorted and deviate from a perfect hexagonal lattice near a corners of the STM image (cf. Fig.~1d of Ref.~\onlinecite{Yazdani2024}), yet a well-defined WC structure remains visible over the rest of the imaged region. Therefore, a ``pinned WC'' does not explain the phenomenon because it becomes simply a tautology for the AS phase unless a compelling explanation is provided why the WC decides to be pinned at $\nu_c \approx 0.093$, and not at a higher (or a lower) filling.

Ref.~\onlinecite{Yazdani2024} clearly shows the existence of WC everywhere except at precise odd denominator filling. We comment that Ref.~\onlinecite{Yazdani2024} does not show any 1/5 FQHL, most likely because the system is not clean enough to manifest the 1/5 FQH gap. 
We also note that the gradual crossover behavior of the transition from the WC phase to the AS phase in Fig.~2 of Ref.~\onlinecite{Yazdani2024} is more consistent with a disorder-induced Anderson localization of the electrons into a spatially random amorphous phase rather than the sudden pinning of a WC by disorder. Most importantly, Fig.~2a and Fig.~2i in Ref.~\onlinecite{Yazdani2024} are manifestly representing a random electron configuration with little evidence for a pinned WC.

The usual argument for the existence of WC at a low filling is grounded in approximate energetics calculations and comparisons between WS and FQHL phases using trial wavefunctions, establishing the stability of the WC phase at some low filling factor (typically $<1/5$) \cite{Maki1983, Lam1984, Levesque1984, Ortiz1993, Zhu1993, Zhu1995, Price1995, Kamilla1997, Yi1998, Yang2001, Mandal2003, Archer2013, Peterson2003}. The serious problems in such numerical calculations are: (1) exclusion of disorder completely; (2) limitations of the trial wavefunction choices for both the WC phase and the FQHL phase. Thus, their reliability in determining $\nu_c$ is highly suspect, because the choice of wavefunctions leads to a bias in the energetic comparison, and more importantly, the neglect of disorder makes any application to the emergence of the disordered AS phase in Ref.~\onlinecite{Yazdani2024} questionable. In addition, Ref.~\onlinecite{Yazdani2024} directly refutes the core argument that the WC phase is stabilized below some critical filling since Ref.~\onlinecite{Yazdani2024} clearly observes a WC everywhere except for $\nu < 0.093$. Thus, the physics of the crossover to AS is surely not about two competing interacting phases (WC versus FQHL) which applies only for $\nu \approx 1/3$ (but not at $\nu \approx 0.093$ where there is no competing nearby FQHL phase in the STM images), but about the role of disorder in creating the AS phase.
The experimentally observed strongly insulating phase at low and nonuniversal filling factors necessarily requires the existence of disorder, and hence the neglect of disorder in the energetic comparison makes the applicability of the energetic comparison to the experimental transport properties questionable.  In addition, Ref.~\onlinecite{Yazdani2024} shows that the WC does not become amorphous until the filling factor decreases to a very low value. This highlights a crucial distinction: although disorder is expected to pin the WC and cause insulating behavior (at sufficiently weak current and low temperature), a pinned crystal is structurally distinct from a disordered amorphous phase.

In the current work, we take a radically alternative approach considering only disorder and neglecting all interaction effects to determine where the disorder-induced crossover to the AS phase should occur in the absence of any interaction.  This is complementary to the standard approaches to the problem~\cite{Maki1983, Lam1984, Levesque1984, Yi1998, Zhu1993, Zhu1995, Yang2001, Mandal2003, Archer2013, Peterson2003, Shayegan1997,Fertig1997} where disorder is completely ignored taking into account only interaction with the assumption that any energetic transition to the WC immediately creates a strongly localized insulator.  The reality is most likely somewhere in between, but obtaining a complete high-field phase diagram in the presence of both disorder and interaction is one of the most difficult tasks in physics.
Our theory below for determining the critical filling factor $\nu_c$ is mostly qualitative, at best semi-quantitative, involving drastic approximations and physical arguments since the microscopic problem including both disorder and interaction on an equal footing allowing for the emergence of all three phases (FQHL, WC, AS) observed in Ref.~\onlinecite{Yazdani2024} is intractable. Since our focus is the AS phase, we mostly discuss the issue of strong localization in the LLL in the presence of disorder. This approach is justified a posteriori by its ability to provide a meaningful explanation for the experimental results on $\nu_c$. We emphasize that we use Ref.~\onlinecite{Yazdani2024} only as an inspiration and a motivation for our work, and we do not by any means are providing a quantitative theory for the observations of Ref.~\onlinecite{Yazdani2024}.


\begin{figure}
    \centering \includegraphics[width=\linewidth]{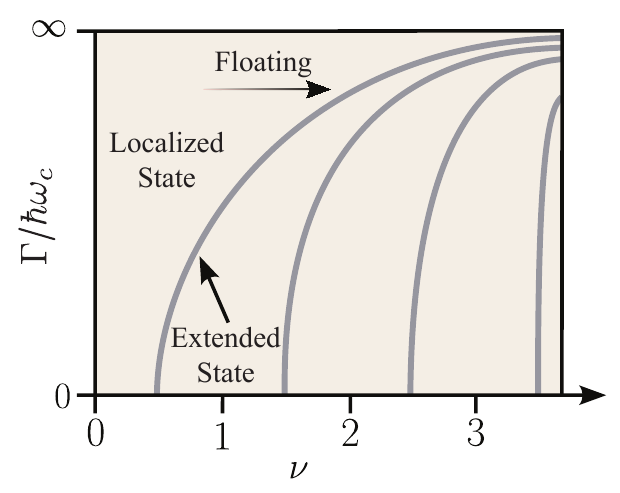}
    \caption{%
    Schematic for the zero temperature phase diagram of the localized and delocalized states in IQHE, demonstrating floating as the disorder broadening increases relative to $\hbar \omega_c$. At low disorder, the Landau levels are located at half integers and float to higher fillings with increasing disorder. Note that we consider a fixed magnetic field keeping $\hbar \omega_c$ constant, and change the filling factor or chemical potential by increasing density. Note also that the floating or the levitation of the individual Landau levels slows down with increasing Landau level number, and the disorder-induced localization due to floating begins at the lowest Landau level, moving up in Landau level as disorder increases.%
    \label{fig:phase_diagram}}
\end{figure}

\section{Determination of $\nu_c$ and heuristics}

As emphasized above, determining the critical filling for the transition to the AS phase from first principles (e.g., quantum Monte Carlo based on a microscopic Hamiltonian including both disorder and electron-electron interactions with LL mixing, allowing for all possible phases: FQHL, WC, AS) is essentially impossible since the problem involves strong disorder physics in a non-perturbative interacting problem in the LLL. In fact, just including disorder in the physics of FQHL is an extremely hard problem, which has rarely been attempted (always in very small systems using simplified Hamiltonians) \cite{Huang2025Thermal,Zhang1985,Rezayi1985}. Using energetics arguments in determining Anderson localization is not a meaningful theoretical approach. Instead, we use several heuristic approaches, which should give a disorder-induced transition at a threshold value of the filling factor, and we discuss their relevance and applicability both to the transition into the AS phase in general and to the specific experimental observations of Ref.~\onlinecite{Yazdani2024}.
Our approach is motivated by our earlier work on the integer quantum Hall effect (IQHE) where we studied in depth the role of disorder, temperature, and LL filling on IQHE plateau in noninteracting 2D systems~\cite{Yi-Thomas2025}.  We now adapt this approach to the LLL to comment on the localization physics at low filling. 

First, we discuss the noninteracting situation in the LLL, which has been studied in great depth recently \cite{Yi-Thomas2025}, where all states are localized for $\nu < 1/2$ and the Hall conductivity is zero for $\nu < 1/2$ with no IQHE. The localization threshold floats up with increasing disorder as shown in the non-interacting phase diagram in Fig.~\ref{fig:phase_diagram} taken from our recent work~\cite{Yi-Thomas2025}. 
Due to the floating-up of extended states when disorder is strong $\Gamma > \hbar \omega_c$, the width of the first $\rho_{xy}=h/e^2$ plateau shrinks more rapidly with disorder than that of the second plateau at $\rho_{xy}=2h/e^2$~\cite{Yi-Thomas2025}, as illustrated in Fig.~\ref{fig:phase_diagram}. Consequently, at finite temperature the first plateau can be completely masked by the neighboring broadened extended state, rendering it unobservable in transport measurements.
This is totally consistent with experiments—in strongly disordered 2D systems one never observes the basic $\rho_{xy} = h/e^2$ quantization because the LLL is completely localized with no extended states. In such a situation, of course, neither the FQHE nor the WC can manifest itself in the LLL since such interaction-induced phases are much more fragile than the non-interacting IQHE in the presence of strong disorder.

Note that this is not an energetic argument, but a more nuanced heuristic involving disorder-induced LL broadening, which, if it becomes comparable to the LL separation, i.e., the cyclotron energy $\hbar \omega_c$, there is no IQHE, FQHE, WC—everything in the LLL is then AS with $\nu_c > 1$. This was in fact the situation for the original discovery of the IQHE by von Klitzing \cite{Klitzing1980}, who did not see any LLL IQHE with a quantized Hall resistance of $h/e^2$ since the disorder was far too strong in the relevant Si MOS 2D samples he studied. Of course, the original IQHE experiment could not possibly discover any FQHE because of the strong disorder with $\nu_c > 1$. The FQHE and the associated FQHL evolved later at $\nu=1/3$ with the improvement in samples associated with the switch from dirtier 2D Si MOS samples (with mobility $\mu \sim 10^3$ cm$^2$/Vs) to cleaner 2D GaAs samples (with $\mu \sim 10^5$ cm$^2$/Vs)~\cite{Tsui1982}. In fact, with this improvement in the quality because of the disorder suppression, $\nu_c$ was pushed below 1/3 (since 1/3 FQHE was observed), and consequently, the IQHE with $\rho_{xy} = e^2/h$ quantization associated with the LLL manifested in this FQHE experiment too \cite{Tsui1982}. The manifestation of the $h/e^2$ LLL IQHE is a necessary (but not sufficient) condition for the manifestation of any LLL FQHE because $\nu_c < 1$ is essential for FQHL to form in the LLL. If the whole LLL is a disorder-induced strongly localized AS, there is no QHE (or WC) at all.

\begin{figure}
    \centering
    \includegraphics[width=0.8\linewidth]{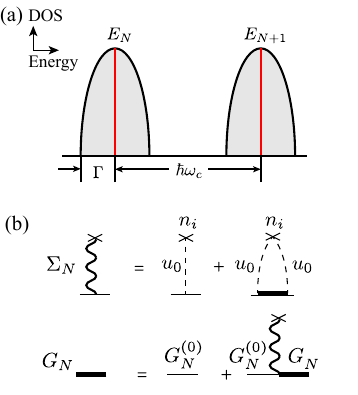}
    \caption{(a) Density of states (DOS) for a 2D electron gas in a magnetic field, calculated within the self-consistent Born approximation (SCBA). 
    The Landau levels, centered at energies $E_N$, exhibit semi-elliptic broadening with a spectral width $\Gamma$ and are separated by the cyclotron energy $\hbar\omega_c$. 
    (b) Diagrammatic representation of the SCBA. 
    The first line illustrates the self-consistency condition, where $\Sigma_N$ is constructed from a single impurity scattering line (dashed line, characterized by concentration $n_i$ and short-range potential $u_0$) dressing the full propagator $G_N$.
    The second line depicts the Dyson equation relating the full Green's function $G_N$ (thick line) to the bare Green's function $G_N^{(0)}$ (thin line) and the self-energy $\Sigma_N$. 
    }
    \label{fig:LL_SCBA}
\end{figure}

We now develop this line of reasoning into a quantitative formula. The basic idea is that the disorder-induced LLL broadening must be smaller than the effective chemical potential (or the Fermi energy) in the LLL for any interaction effects to manifest. If not, the system is an AS. Therefore, by equating the disorder broadening of the LLL to the effective LLL chemical potential, we get the condition defining the critical filling factor for strong localization. This leads to the following equation:
\begin{equation}
    \Gamma_{\text{LLL}} = \nu_c \hbar \omega_c. \label{eq:gamma_LLL}
\end{equation}
Here, $\Gamma_{\text{LLL}}$ is the LLL broadening, which depends on the disorder strength and the applied magnetic field. A microscopic calculation of the broadening in the presence of both disorder and interaction is essentially impossible and has never been attempted. However, we can make drastic approximations to make progress. We can assume that the disorder is short-ranged (perhaps because of screening) and then use the self-consistent Born approximation (SCBA) to solve the multiple impurity scattering diagrams shown in Fig.~\ref{fig:LL_SCBA}~\cite{Ando:1974}. Using SCBA, the integral equation of Fig.~\ref{fig:LL_SCBA} can be solved leading to the following formula for the LL density of states (DOS) $D(E)$ and the broadening $\Gamma$ (see Appendix for derivation):
\begin{align}
    D(E) = \frac{1}{2\pi l_B^2} \sum_{N} \frac{2}{\pi \Gamma}\qty[1 - \qty(\frac{E-E_N}{\Gamma})^2]^{1/2}, \label{eq:dos_scba}
\end{align}
\begin{align}
    \Gamma^2 = \frac{2}{\pi} \hbar \omega_c \frac{\hbar}{\tau}. \label{eq:gamma_scba}
\end{align}
Here, $l_B = (\hbar c/eB)^{1/2}$ is the magnetic length (or the cyclotron radius in the LLL), and $\tau$ the zero-field disorder scattering time.. $E_N = (N+1/2)\hbar \omega_c$, $N=0,1,2,3,\dots$, is the energy of the $N$-th LL (with $N=0$ being the LLL).
Equation~\eqref{eq:gamma_scba} can be intuitively understood as follows.
By forming LLs in strong $B$ field, the DOS $D(E)$ of a LL with a width $\Gamma$ increases compared to the parent 2D DOS $D_0(E)$ at $B=0$, following $D = D_0 \hbar \omega_c/\Gamma$.
Meanwhile, because the DOS increases, the scattering also becomes stronger, and the disorder broadening $\Gamma$ at $B\neq 0$ increases compared to the disorder broadening $\Gamma_0 = \hbar/\tau$ at $B=0$, following $\Gamma = \Gamma_0 D/D_0$.
By solving the two equations $D = D_0 \hbar \omega_c/\Gamma$ and $\Gamma = \Gamma_0 D/D_0$ self-consistently, we obtain $\Gamma^2 = \hbar \omega_c \Gamma_0$, which is the same as Eq. \eqref{eq:gamma_scba} up to a numerical factor.
In Eqs.~\eqref{eq:dos_scba} and \eqref{eq:gamma_scba}, the magnetic field is held fixed, so that $\hbar\omega_c$ remains constant, and the filling factor is tuned solely by increasing the carrier density—precisely the same strategy used to control $\nu$ in Ref.~\onlinecite{Yazdani2024}.
The quantity $\tau$ in Eq.~\eqref{eq:gamma_scba} is simply the corresponding zero-field disorder scattering time, which, in principle, should be the ``quantum'' scattering time (also called the single-particle scattering time), and not the transport relaxation time \cite{DasSarma1985}. The approximations of 1) restricting to the LLL, 2) neglect of LL mixing, and 3) using short-range disorder together considerably simplify the problem to an analytical formula for $\nu_c$ which is given by combining Eqs.~(\ref{eq:gamma_LLL}) and \eqref{eq:gamma_scba}:
\begin{equation}
    \nu_c = \frac{\Gamma}{\hbar\omega_c} \label{eq:nuc0}
\end{equation}
with
\begin{equation}
    \Gamma^2 = \frac{4}{\pi} \hbar \omega_c \Gamma_0 \label{eq:gamma2}
\end{equation}
where
\begin{equation}
    \Gamma_0 = \frac{\hbar}{2\tau}. \label{eq:gamma0}
\end{equation}
Combining Eqs.~(\ref{eq:nuc0})-(\ref{eq:gamma0}), we get a deceptively simple analytical form for $\nu_c$ defining the threshold for the emergence of the AS phase (i.e., $\nu_c = 0.093$ in Ref.~\onlinecite{Yazdani2024}):
\begin{equation}
    \nu_c \sim \qty(\frac{\Gamma_0}{\hbar\omega_c})^{1/2} \label{eq:nuc}
\end{equation}
In writing Eq.~(\ref{eq:nuc}), we have simply ignored the factor of $4/\pi$ in Eq.~(\ref{eq:gamma2}) since $4/\pi \sim O(1)$, and the theory is not precise enough to carry numerical prefactors of $O(1)$.

Before proceeding any further, we must test how well the simple Eq.~(\ref{eq:gamma_LLL}) stands up to reality by checking against the extreme cases of extremely disordered Si MOS~\cite{Klitzing1980} and the very clean BLG \cite{Yazdani2024}. In the Si MOS dirty 2D system, $\mu \sim 10^2$ cm$^2$/Vs, implying that $\Gamma_0 \sim 30$ meV assuming the appropriate Si 100 effective mass $m\sim 0.19 m_e$. Then, for a magnetic field $B \sim 10$ T, the cyclotron energy $\hbar \omega_c \sim 6$ meV, giving:
\begin{equation}
    \nu_c \sim \sqrt{5} > 2. \label{eq:8}
\end{equation}
We note, however, that if the mobility is 10 times higher (a still modest $\sim 10^3$ cm$^2$/Vs) and the $B$-field is a factor of 2 higher ($\sim 20$ T), then $\nu_c$ decreases by a factor of 4, still remaining above 1/2, implying no IQHE or FQHE in the LLL in Si MOS systems.

Turning to the clean BLG of Ref.~\onlinecite{Yazdani2024}, which motivated our work, we use $\mu \sim 10^6$ cm$^2$/Vs consistent with the high quality of the sample (corresponding to a long mean free path $> 0.1$ micron) and use the BLG effective mass $\sim 0.04m_e$ to obtain (at $B=10$ T):
\begin{equation}
    \nu_c \sim 0.023. \label{eq:nuc_BLG}
\end{equation}
This critical filling is smaller than the experimental value of $0.093$, but given the simplicity of our theory and many approximations, perhaps this is an acceptable answer under the circumstances. We note that the actual LL broadening is expected to be larger than our estimated values because of our approximations of short range disorder, neglect of LL mixing, and using the transport lifetime to estimate $\Gamma$ (instead of the correct quantum broadening which is always larger than the transport broadening). In particular, all three approximations tend to reduce $\Gamma$, so our estimated $\nu_c$ is a lower bound for the critical filling factor to enter the strongly localized AS phase.

In the original 2D GaAs system for the discovery of the FQHE~\cite{Tsui1982}, $\mu=10^5$ cm$^2$/Vs and $m=0.07m_e$ with $B \sim 20$ T. This leads to:
\begin{equation}
    \nu_c \sim 0.05 \sim 1/20. \label{eq:10}
\end{equation}
This is again smaller than the experimental $\nu \sim 0.2$ where the original FQHE GaAs sample goes into an insulating phase. Comparing with other high-mobility GaAs samples, we find that the discrepancy between our estimated $\nu_c$ and the experimental filling factor, where the system enters into the strong insulator phase in transport, persists in most samples with our estimated $\nu_c$ always being typically smaller than the experimental filling factor for the LLL to become insulating. 
One possible explanation for this discrepancy is that the experimentally observed insulating phase at $\nu \sim 0.2$ is a pinned WC instead of AS; in this scenario, the predicted AS would emerge at a even lower filling closer to our predicted $\nu_c \sim 0.05$ (this scenario is similar to the case in Ref.~\onlinecite{Yazdani2024}). 
Alternatively, the discrepancy may arise from the nature of modulation doping in GaAs samples, and hence the transport broadening could be orders of magnitude smaller than the quantum broadening since most of the scattering by the remote dopants is forward scattering which does not contribute to transport broadening \cite{DasSarma1985}. It would be quite interesting in this context to measure both the transport and the single particle lifetime in a 2D GaAs sample to see if the discrepancy between theory and experiment is resolved by the enhancement of the quantum broadening over the transport broadening. It is encouraging that the theoretical $\nu_c$ is a lower bound since the experimental $\nu_c$ always seems larger than the theoretical estimate.

Before concluding this section, we emphasize our qualitative findings:
\begin{align}
    \nu_c &\sim (\Gamma_0 / \hbar\omega_c)^{1/2} \label{eq:nuc_gamma0_omegac} \\
    &\sim \frac{\Gamma_0 m}{\hbar^2 n} \sim \frac{\Gamma_0}{E_{F0}}. \label{eq:nuc_gamma0_n}
\end{align}
Here $n$ is the 2D carrier density, $E_{F0} \sim \hbar^2 n/m$ is the Fermi energy at $B=0$, and we have replaced the $B$-dependence by the $n$-dependence simply by noting that the 2D LL filling factor is given by:
\begin{equation}
    \nu = 2 \pi n l_B^2. \label{eq:nu_nlb2}
\end{equation}
We mention that Eqs.~(\ref{eq:nuc_gamma0_omegac}) and (\ref{eq:nuc_gamma0_n}) are appropriate respectively for situations involving fixed magnetic field (with density being varied to tune the filling factor) as in gated samples (e.g., Si MOS, BLG) and involving fixed density (with magnetic field being varied to tune $\nu$) as in modulation doped GaAs samples. A direct experimental verification of the scaling on sample quality, magnetic field or density as in Eqs.~(\ref{eq:nuc_gamma0_omegac}) and (\ref{eq:nuc_gamma0_n}) will go a long way in further understanding the nature of the AS phase.

There are alternative heuristic methods for deriving analytical expressions for $\nu_c$ that prioritize different physical mechanisms, contrasting with our focus on the simple equality between the chemical potential and disorder broadening as the criterion for the AS threshold. While our considerations focus primarily on the role of disorder in inducing strong electron localization, electron-electron interactions naturally favor a WC phase, as evidenced in Fig.~2 of Ref.~\onlinecite{Yazdani2024} for $0.2 < \nu < 0.31$. The question naturally arises what happens to this WC phase in the presence of disorder, which is obviously present in the sample for low enough carrier density, as can be seen for the amorphous structure at $\nu \approx 0.093$. 
The standard expectation is that disorder pins the WC immediately, rendering it insulating. However, transport experiments in high-quality n-GaAs samples do not observe a strongly insulating phase for $\nu > 0.2$. This creates an apparent tension with the results in Ref.~\onlinecite{Yazdani2024}, where a WC is observed across nearly all fillings $0.334>\nu>\nu_c\sim0.093$ except a small range near $\nu \approx 0.334$.
In contrast, transport studies on p-GaAs hole systems \cite{Wang:2025b} report insulating behavior extending up to $\nu=1/3$, qualitatively similar to observations in BLG. This discrepancy can be attributed to LL mixing: GaAs holes possess a larger effective mass, and BLG has a significantly lower dielectric constant ($\kappa \approx 4$ in hBN versus $\kappa \approx 13$ in GaAs), both of which enhance LL mixing. Taken together, these observations suggest that enhanced LL mixing shifts the critical filling for the onset of the insulating phase toward higher values.
Within this framework, the insulating phase first appears as a pinned WC. As the filling or carrier density decreases, disorder eventually destroys the crystal structure, transitioning the system into an AS at $\nu < \nu_c$.
This interpretation is consistent with Ref.~\cite{Yazdani2024}, where the onset of the pinned WC ($\nu \sim 0.4$) occurs at a significantly higher filling than the transition to the AS phase ($\nu \sim \nu_c = 0.093$).
Nevertheless, how $\nu_c$ behaves in the presence of both disorder and interaction remains an open question. There is obviously no easy or sharp answer to this question as the disorder-interaction interplay is complex and highly sample dependent. 

We emphasize here that our formula for disorder and field (or density)-dependent $\nu_c$ does not incorporate interaction effects, and as such, it is independent of the competing phase being FQHL or WC (or something else). The theory asserts that once $\nu < \nu_c$ the system is a randomly spatially localized insulator preempting any other phase existing at that $\nu$ value in the absence of disorder. So, one clear prediction based on our theory is that a gated BLG sample, which is roughly 10 times more disordered, would have a $\nu_c$ which is 10 times higher if only the density is being tuned to vary the filling keeping the magnetic field constant. This implies that a sample with 10 times the disorder of Ref.~\onlinecite{Yazdani2024} would make a transition into the AS phase at $\nu_c \sim 0.93$ if the same magnetic field is being used in the experiment. Therefore, a much dirtier (10 times) sample would manifest no WC or FQHL phase in the LLL, it will only manifest a spatially random quenched electronic AS phase. It is possible that our theory actually becomes more accurate with increasing disorder in which case the transition would happen at 10 times our theoretically estimated $\nu_c$ for the current sample of Ref.~\onlinecite{Yazdani2024}, which would be $\nu_c \sim 0.5$. Then, there would be a transition from a WC phase to the AS phase at some rather high filling around $\nu \sim 0.5$, so all FQHL phases below $\nu \sim 0.5$ will be suppressed by disorder. This leads to a striking and readily testable prediction: systematically increasing sample disorder should directly reveal this shift in the critical filling.

\section{Impurity density Consideration}

Our SCBA treatment of disorder broadening, which yields the critical filling factors in Eqs.~\eqref{eq:nuc_gamma0_omegac} and \eqref{eq:nuc_gamma0_n}, explicitly assumes short-range disorder. In this leading-order theory, the zero-field broadening is directly proportional to the impurity density $n_i$:
\begin{equation}
\Gamma_0 \sim \frac{\hbar^2 n_i}{m}. \label{eq:gamma0ni}
\end{equation}
It turns out that even Coulomb scattering at zero field becomes short-ranged in 2D for low carrier density $n$ such that $k_F \ll q_{TF}$, where $k_F$ and $q_{TF}$ are respectively the zero-field Fermi momentum and Thomas-Fermi screening wavevector defined by:
\begin{equation*}
    k_F = (4\pi n)^{1/2}; \quad q_{TF} = \frac{me^2}{\kappa \hbar^2} = \frac{1}{a_B},\label{eq:kF_qTF}
\end{equation*}
where $a_B$ is the effective Bohr radius and $\kappa$ is the lattice dielectric constant for the semiconductor.
The Fourier transformation of the screened Coulomb disorder potential is then given by:
\begin{equation*}
    u_i (q \ll q_{TF}) = \frac{2\pi e^2}{q + q_{TF}} \approx \frac{2\pi e^2}{q_{TF}},
\end{equation*}
which is independent of the carrier density and momentum transfer $q$, reflecting the energy independence of the 2D density of states. Thus, the 2D screened Coulomb disorder is effectively short-ranged. Using the Born approximation to calculate the impurity scattering induced zero-field broadening then gives:
\begin{equation*}
    \Gamma_0 = \left( \frac{m n_i}{4\pi \hbar^2} \right) \left( \frac{2\pi e^2}{\kappa q_{TF}} \right)^2 \sim \frac{\hbar^2 n_i}{m}.
\end{equation*}
The Ioffe-Regel-Mott (IRM) criterion~\cite{ioffe1960,Mott:1969} for the zero-field 2D metal-insulator transition (MIT) is $\Gamma_0 = E_{F0} = \hbar^2 k_F^2/2m$.
which then immediately leads to the simple condition for the Anderson localization that the carrier density equals the charged impurity density:
\begin{equation}
    n = n_i. \label{eq:nmit}
\end{equation}
In the zero-field case also, transport experiments often interpret the manifestation of the 2D MIT (often happening at $n\sim n_i$) as the Fermi liquid to the WC transition. Therefore, a question naturally arises what Eq.~(\ref{eq:nmit}) would imply in the strong-field LLL situation, particularly since our heuristic criterion for the LLL localization, as defined by Eq.~\eqref{eq:gamma_LLL}, is simply the modified IRM criterion $\Gamma_\nu = E_{F\nu} = \nu \omega_c$ with $\Gamma_\nu$ and $E_{F\nu}$ being respectively the filling factor dependent disorder broadening and the chemical potential.

We now investigate the implications of Eq.~(\ref{eq:nmit}) as the strong localization criterion in the strong-field situation. We notice a problem right away if the 2D system is tuned by a varying magnetic field $B$ to vary the filling instead of varying the carrier density. (This problem is absent in the zero field case where the carrier density is the only tunable parameter to produce the MIT in a given sample.) In a given sample with fixed $n$, either $n > n_i$ (``strong localization'') or not (``no localization''). We first ignore this problem, and assume that $B$ is fixed and the tuning parameter is $n$ with $2\pi l_B^2 n = \nu$ fixing the necessary $B$ field. Using the definition of the magnetic length, $l_B^2 = \hbar c/eB$ and $\omega_c = eB/mc$, it then immediately follows that Eq.~(\ref{eq:nmit}) gives the following critical filling factor $\nu_c$ for the localized insulating state:
\begin{equation}
    \nu_c = \frac{8\pi^3 \hbar n_i}{m \omega_c}. \label{eq:nuc_nmit}
\end{equation}
At first, $\nu_c$ of Eq.~(\ref{eq:nuc_nmit}) looks very different from that in Eq.~(5). But we can put them in similar forms by realizing that the zero-field broadening $\Gamma_0 \sim n_i$, leading to the following scaling form for $\nu_c$ given by Eq.~(\ref{eq:nuc_nmit}):
\begin{equation}
    \nu_c \sim \frac{\Gamma_0}{\hbar \omega_c}. \label{eq:nuc_gamma0_short}
\end{equation}
This scaling is linear instead of being a square root in $\Gamma_0/\hbar\omega_c$ as in Eq.~\eqref{eq:nuc}, but they have the same qualitative dependence: both predict a critical filling increasing with increasing (decreasing) zero-field broadening (cyclotron frequency). This is indeed the qualitative behavior of the observed $\nu_c$ in the transition to the strongly insulating state at low LLL filling, which is accepted uncritically as the signature for a transition to a strong-field WC. In our non-interacting theory, however, the transition is to a strongly localized Anderson insulator. 

As a sanity check, we have estimated some numbers for various systems using Eq.~(\ref{eq:nuc_nmit}) for the critical filling. We get the following results. For Si MOSFETs, which are generally highly disordered with $n_i > 10^{11}$ cm$^{-2}$ we get: $\nu_c = 4$ and $0.4$ for $n_i = 10^{12}$ and $10^{11}$ cm$^{-2}$ respectively ($B=10$ T in both cases). For 2D n-GaAs high-quality low disorder modulation doped structures, the typical impurity density is low, so we get: $\nu_c = 0.03$ for $n_i = 10^{10}$ cm$^{-2}$ also at $B=10$ T. The agreement with experiments is better for high-disorder Si 2D samples, where typically all samples manifestly strongly insulating state in the LLL, the theoretical critical filling for the low-disorder GaAs sample is quantitatively poor since typically the highly insulating phase manifests itself for a filling $< 1/10$ even in the best quality samples. Given the simplicity of our theory ignoring all interaction effects, the factor of 3 disagreement in the predicted $\nu_c$ is not surprising.

We mention that for a classical WC (which may apply to very high magnetic fields with extremely small cyclotron radius, much smaller than the WC lattice constant), so that the electrons can be treated as point particles in zero field, an alternate formula can be derived for $\nu_c$ in the presence of random charged impurities in the 2D electron layer. The derivation is based on earlier work, where the stability of the WC or 2D charge density waves was derived in the presence of its coupling to impurities~\cite{Fukuyama1978}, leading to the stability condition (for 2D screened charged impurities):
\begin{align}
    n > n_i^{1/2}/a_B,
\end{align}
where $n_i$ and $n$ are the charged impurity density and the carrier density, respectively, and $a_B=\kappa \hbar^2/m e^2$ is the effective Bohr radius for the 2D semiconductor.  Note that if the impurities are located at a distance $d >a_B$ from the 2D electrons, then $a_B$ here is to be replaced by $d$.  Converting this equation to the condition for $\nu_c$ by using $\nu=2\pi l_B^2 n$, we get:
\begin{align}
    \nu_c \sim \frac{(\Gamma_0 \,\mathrm{Ry})^{1/2}}{\hbar\omega_c},
\end{align}
where $\Gamma_0$ is the usual zero-field level broadening [cf. Eq.~\eqref{eq:gamma0ni}], and $\mathrm{Ry}\sim e^2/(\kappa a_B)\sim \hbar^2/(m a_B^2)$ is the effective Rydberg energy.   
Note that this scaling, although differing in the detailed power laws from the ones derived above,  still predicts that the critical filling for the crossover to a strongly localized and fragmented AS should increase with increasing impurity broadening and decrease with increasing magnetic field. The difference is that the transition here involves the destruction of the WC phase by the strong coupling to random charged impurities.


\begin{figure}[t]
    \centering
    \includegraphics[width=\linewidth]{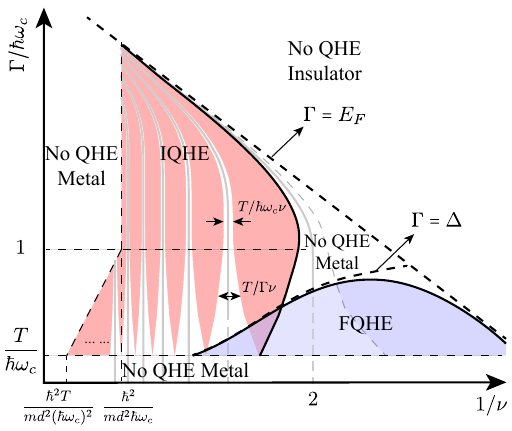}
    \caption{The schematic finite-temperature ($T\ll \hbar \omega_c$) phase diagram (in log-log scale) is presented in the plane of disorder strength versus inverse filling factor. The regions shaded in red and blue correspond to the integer and fractional quantum Hall insulator phases, respectively, where $\sigma_{xx}=0$ and $\sigma_{xy}/(e^2/h)$ is an integer and fraction, respectively. The upper dashed line marks the disorder-induced metal-insulator transition, defined by $\Gamma=E_F$. Below this upper dashed line, the white region represents a metallic phase without quantized Hall conductivity plateaus. Above the dashed line, the white region corresponds to an insulating phase where both $\sigma_{xx}$ and $\sigma_{xy}$ vanish. The gray curves within the quantum Hall regime indicate the delocalized states at the center of each Landau level, which are broadened by finite temperature, extending into the metallic phase. 
    }
    \label{fig:phase}
\end{figure}

\begin{figure}[t]
    \centering
    \includegraphics[width=\linewidth]{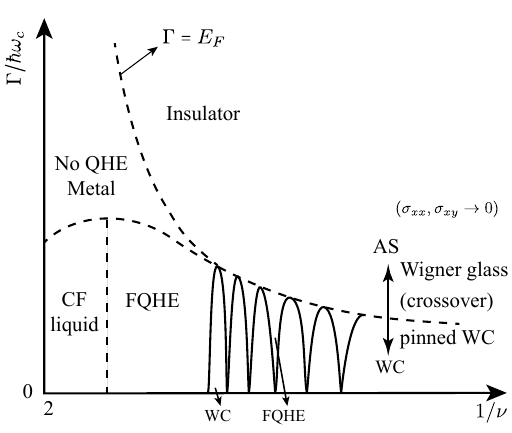}
    \caption{The schematic $T=0$ phase diagram for the lowest LL regime ($1/\nu >2$) is presented in the plane of disorder strength versus inverse filling factor. The upper dashed line marks the disorder-induced metal-insulator transition, defined by $\Gamma=E_F$. At relatively small disorder and close to $\nu=1/2$, the effective magnetic field for the 2-flux composite fermions is small, and we have delocalized composite fermi liquid (CFL). CFL can be destroyed by disorder into electron fermi liquid and eventually to Anderson insulator when disorder is sufficiently strong. Moving away from $\nu=1/2$, the CFL transitions into FQH states, analogous to the metal-to-IQHE transitions shown in Fig.~\ref{fig:phase}. At low filling factors (large $1/\nu$), a pinned WC emerges, confining FQH states to narrow ranges around fractional fillings. With finite disorder and decreasing $\nu$, the pinned WC eventually crosses over into Wigner glass and Anderson solid phases.}
    \label{fig:phase_LLL}
\end{figure}

\section{Phase Diagram}
We next discuss the qualitative phase diagram implied by the above considerations, now putting in empirically and phenomenologically the known 2D strong-field interaction effects. 
Extensive research has explored the LLL phase diagram in the limit of zero disorder, with several experimental studies claiming quantitative agreement with these "clean" theoretical models \cite{Deng2016, Ma2020, Shayegan2021}. However, these theoretical treatments often neglect disorder entirely, comparing ground-state energies only for WC and FQHL wavefunctions.
In contrast, experimental phase diagrams are inherently sensitive to disorder. The standard interpretation identifies the onset of a strongly resistive state at low filling as the WC phase overcoming the FQH phase.
However, this interpretation often overlooks the possibility of a disorder-induced AS. Crucially, both a pinned WC and an AS can exhibit strongly insulating behavior, making transport data alone insufficient to distinguish between them.

Combining the existing theoretical comparisons between WC and FQHL energetics with our disorder results, we propose a schematic phase diagram shown in Figs.~\ref{fig:phase} and \ref{fig:phase_LLL}, each highlighting schematically and qualitatively different aspects of the competing phases arising from the existence of disorder as well as WC, IQHE, FQHE, which are all distinct phases arising from different physics. The $\Gamma=E_F$ line is the key disorder contribution emphasizing the possible disorder-induced insulating AS phase.

The broad phase diagram (in color) shown in Fig.~\ref{fig:phase} is a high-level construct combining our earlier work on IQHE \cite{Yi-Thomas2025} with the current work. It shows the broad features arising from IQHE and FQHE coupled with disorder, with the FQHE part being associated with interactions. 
For weak disorder ($\Gamma < T$), thermal activation delocalizes all states originally localized at $T=0$ (shaded red in Fig.~\ref{fig:phase}) and completely suppresses the QHE, so there is no plateau in $\sigma_{xy}=\nu e^2/h$. 
In the intermediate disorder regime ($T < \Gamma < \hbar \omega_c$), the broadening of the extended states in the filling factor follows $T / \Gamma$. Because Fig.~\ref{fig:phase} employs the inverse filling factor as the horizontal axis, where the interval between consecutive integer fillings $\nu$ and $\nu+1$ scales as $\sim1/\nu$, the corresponding broadening in this inverse filling factor axis follows $T / (\Gamma \nu)$.
When the disorder strength increases to $\hbar \omega_c < \Gamma < N \hbar \omega_c \sim E_F$, this broadening in filling factor saturates at $T / \hbar \omega_c$ [or $T / (\hbar \omega_c\nu)$ in the inverse filling factor as shown in Fig.~\ref{fig:phase}]. 
For strong disorder ($\Gamma > N \hbar \omega_c$), the extended states shift to higher filling factors, which is known as the floating physics. 
At sufficiently large filling factors, where the cyclotron radius $R_c$ is larger than the disorder potential correlation length $d$ so that $R_c / d > \max(1, \hbar \omega_c / \Gamma)$, the diffusion of electron guiding centers becomes efficient, disrupting semiclassical localization in the quantum Hall regime~\cite{Yi-Thomas2025}. 
As a result, the system undergoes a crossover from IQHE phase to the no-QHE metallic phase where $\sigma_{xy} \propto 1/B$ and $\sigma_{xx}$ constant in $B$ follows the conventional Drude formula. 
If $T>\hbar \omega_c$, all QHE phases are destroyed and leading to the semiclassical metallic phase without QHE plateau in $\sigma_{xy}$.

In the FQHE phase, electron-electron interactions introduce an energy scale, the many-body gap $\Delta$, that competes with $\Gamma$, $T$, and $E_F$, as shown in Fig.~\ref{fig:phase}. 
FQHE is destroyed and crosses over to a metallic state without quantized Hall conductivity plateau, if $\Gamma>\Delta$ or $T>\Delta$.
In principle, $\Delta(\nu)$ is a nontrivial function of $\nu$, but it is well-known that $\Delta(\nu)$ decreases with decreasing $\nu$ (and FQHE becomes increasingly scarce at lower LLL filling). For sufficiently low temperature and disorder $T,\Gamma<\Delta$, experiments indicate that the FQHE persists up to the second Landau level, with the most stable states with larger $\Delta(\nu)$ occurring in the lowest LL at filling factors $1/\nu > 2$. If we assume that $\Delta(\nu)$ takes a dome shape centered somewhere in the lowest LL, e.g. $\nu=1/3$, then by lowering disorder with better sample quality, the $\nu=1/3$ FQHE plateau would first develop when $\Gamma=\Delta(\nu=1/3)$, and FQHE at other fractional fillings would develop when disorder is lowered further. Using this assumption for $\Delta(\nu)$, FQHE at higher LL only shows up for very clean sample since $\Delta(\nu)$ is very small for FQHE at higher LL, which is what experiments show. This assumption agrees with the experimental observation that more plateaus are developed as disorder is lowered. The existence of FQHE plateau is due to localization of quasielectrons and quasiholes near the fractional filling~\cite{Stormer:1999}. Additionally, for weak disorder where $\Gamma<T$, we suspect the FQHE melts into a metallic phase without QHE, analogous to the IQHE case, because of the thermal broadening of extended states of FQHE plateau transitions~\cite{Jain_Shayegan:1990}. Consequently, the FQHE phase exists only within the disorder range $T<\Gamma<\Delta$. 
This tight constraint makes FQHE a fragile phenomenon--it is suppressed by disorder and temperature.

The most relevant phase diagram for the current work is in Fig.~\ref{fig:phase_LLL}, which restricts to the LLL, and shows qualitatively all the relevant phases: FQHL, WC, AS. Below the $\Gamma = E_F$ line away from $\nu=1/2$, regime in between two FQH states is generically occupied by WC (or a similar interacting phase, e.g., stripes, with no topological order and no FQHE), as shown in Figs.~2 and 3 of Ref.~\onlinecite{Yazdani2024}. Figure~\ref{fig:phase_LLL} conjectures that in the limit of zero disorder, the FQHE is restricted to a set of measure zero occurring precisely at odd-denominator filling factors. Both WC and FQHL are truncated by the disorder line, and if the disorder is strong, then neither would manifest itself (which is the situation in all samples prior to the 1982 discovery of FQHE). If the disorder is weak, however, many FQHE at many odd denominators would appear, which is the current situation in the very best samples (both GaAs and graphene) where FQHL exists experimentally to denominators as large as 13~\cite{Pan2002,Wang2025}. But eventually, the disorder line comes down at low filling, and may touch the WC/FQHL sequence, terminating their existence with a crossover to the AS phase. Details would depend not only on the disorder strength, but also on the temperature. 
Beyond the general observation that high-quality samples exhibit numerous FQH states while lower-quality ones show few or none (with Ref.~\onlinecite{Yazdani2024} manifesting only the $1/3$ and $2/5$ states) little else can be stated generically. Note that the essence of this phase diagram is that the WC in the presence of disorder goes continuously from a pinned WC to a completely random and amorphous AS with little spatial order depending on both the filling factor and the disorder strength.


\section{Conclusion}

We argue that the extensively studied (in magneto-transport) interplay, competition, and transition between FQHL and WC in the LLL of 2D electron systems is fundamentally affected by the disorder content in the 2D sample. It is a nonuniversal disorder-dependent crossover phenomenon, not easily explicable based just on theoretical energetic comparisons between calculated FQHL and WC phases as is usually done.

In particular, before the discovery of the FQHE in 1982, the LLL was experimentally always strongly insulating by virtue of the low sample quality or high disorder. The 1982 discovery of the FQHE at $\nu=1/3$ in a 2D modulation-doped GaAs sample with mobility $\sim 10^5$ cm$^2$/Vs ($\Gamma \sim 0.1$ meV) found the system going insulating right below 1/3 filling, which could be construed as the WC formation at $\nu < 1/3$ just below the FQHL formation precisely at $\nu=1/3$~\cite{Tsui1982}. Later experiments in better samples, mobility $\sim 6.5 \times 10^7$ cm$^2$/Vs (with transport $\Gamma < 0.01$ meV, although modulation doping could substantially enhance the corresponding quantum single-particle broadening), led to the insulating state moving down to $\nu \sim 1/5$ just around the appearance of a FQHE at $\nu=1/5$. This was emphatically claimed to be the emergence of a WC, using the justification of theoretical pristine energetic calculations showing an FQHL to WC transition at $\nu \sim 1/5$~\cite{Jiang1990,Goldman1990}. More recently, still better samples started showing the existence of FQHE at $\nu \sim 1/7$~\cite{Goldman1988,Shayegan_1_7:2022}, and hints of FQHE down to $\nu \sim 1/9$ or even $1/11$. In this low filling regime, these FQH states typically manifest only as shallow minima superimposed on a dominant, strongly insulating background localized by disorder.

Thus, the indisputable experimental fact is that the transition to the strongly insulating state has a strong disorder dependence, with the observed $\nu_c$, the lower bound for identifying FQH states via longitudinal resistivity ($\rho_{xx}$) minima, shifts toward lower filling factors with decreasing disorder. 
Although a quantitative (or even semi-quantitative) comparison with experiments is impossible even in our simple theory, since the quantum broadening could be much larger than the mobility broadening, the experimental results over a 45-year period (1980-2025) have consistently found a decreasing $\nu_c$ with increasing quality, with the strongly insulating state appearing from $\nu_c > 1$ in the pre-1982 high-disorder situation~\cite{Klitzing1980,Tsui1980} through $<1/3$ in 1982~\cite{Tsui1982} and $\sim 2/7$~\cite{Stormer1983}, all the way to $\sim 1/5$~\cite{Jiang1990,Goldman1990}, and $\sim 1/7$~\cite{Goldman1988,Shayegan_1_7:2022,Wang2025}.

This decreasing $\nu_c$ with decreasing disorder in transport experiments, combined with the STM finding in Ref.~\onlinecite{Yazdani2024} that the 2D system reflects the WC formation generically at all $\nu$ except for some of the precise odd-denominator fillings where FQH states locate, then becoming an amorphous random AS (i.e., a strongly pinned fragmented WC which is equivalent to an AS) at a very low $\nu_c$, leads to our conclusion: Disorder is the driving mechanism for the existence of a critical filling causing a crossover to a strongly insulating phase in the LLL.

Lower the disorder, lower is the $\nu_c$, which is difficult to explain simply based on an energetic comparison between the FQHL and the WC phase, as is always done in the context of the existence of a $\nu_c$ in the transport experiments. Such a disorder-free energy comparison can only produce a unique $\nu_c$ where the two energy curves cross, and is incapable of providing an explanation for the sample-dependence of the observed experimental $\nu_c$. It is entirely possible that the system is always a WC at all $\nu$ except at a set of measure zero of odd denominator fillings where the FQHL forms cusps coming below the WC energy. 
In fact, this is precisely the scenario found in the latest and the best numerical calculations where the pristine system manifests FQHE down to very low odd denominator values of the LLL filling, implying that the FQHL is energetically more favorable than the WC perhaps at arbitrarily low odd denominator fillings in the LLL~\cite{Zuo2020}. 

This impressive recent theoretical study utilizing variational Monte Carlo, density function renormalization group, and exact diagonalization has clarified the energetics of the pristine system, effectively superseding earlier comparisons between FQHL and WC phases~\cite{Zuo2020}. This work demonstrates that in the clean limit, the FQHL remains the ground state at odd-denominator fillings down to $\nu=1/9$ and potentially $\nu=1/11$, possessing lower energy than the competing crystalline phases. Similar result is also found in Ref.~\cite{Peterson2003}. Consequently, the strongly insulating behavior experimentally observed for $\nu \lesssim 1/7$ cannot be ascribed to an intrinsic termination of the FQH series in the clean limit. Instead, these findings provide robust support for the perspective that disorder is the decisive factor in this regime, driving the formation of the insulating state.
Note that our work ignores interaction completely, and finds that the transition to the strongly insulating AS phase occurs at lower $\nu_c$ for higher disorder as observed experimentally, but our agreement with the experimentally observed $\nu_c$ in the high-quality samples is modest, which is understandable given the simplicity of our model and approximations.

Specifically, recent experiments on GaAs~\cite{Wang2025,Wang:2025b} are consistent with the persistence of the FQHL down to ultralow fillings. These studies reveal local stability at precise rational denominators, such as $\nu=1/9$, and extend even to unexpected even-denominator fractions like $\nu=1/6$ and $1/8$. However, these developing states manifest as fragile resistance minima superimposed upon a dominant, highly insulating background where $R_{xx}$ diverges, reaching values as high as $35$~M$\Omega$. Similar data was reported earlier in n-GaAs~\cite{Pan2002}, where FQH signatures from $\nu = 2/11$ down to $\nu=1/7$ appear as $R_{xx}$ minima atop an insulating background. This phenomenology aligns with a percolation model within a strongly localized Anderson insulator. Here, the FQHL forms rare, disconnected puddles that lack long-range connectivity due to the percolation landscape of lakes and mountains~\cite{shklovskii1984}. Consequently, the FQHL contributes to transport only at elevated temperatures, where the partial melting of the insulating background (AS phase at low filling or pinned WC at intermediate filling) allows the characteristic FQH $R_{xx}$ minima to show up.

If one interprets the strongly insulating LLL state to be simply a pinned WC, then the fact that the pinning happens only at a sample-dependent $\nu_c$ remains inexplicable because all samples always have disorder and any emergent WC is always pinned at all fillings in any real sample. The data presented in Fig.~2 of Ref.~\onlinecite{Yazdani2024} is consistent with our interpretation that there is a critical amount of disorder necessary to cause localization into the amorphous phase. The AS phase can be construed to be a highly fragmented strongly impurity-pinned WC phase with no spatial order. Either way, the physics of disorder-induced localization is crucial to understand the sample-dependent LLL crossover to an insulating phase \cite{DasSarma1996Chapter1}.

Our prediction of $\nu_c \sim (\Gamma_0/\hbar \omega_c)^{1/2}$ or $(\Gamma_0 m/\hbar^2 n)$--the two formulae are equivalent since $\nu=2 \pi l_B^2 n$--is in principle verifiable, but in general the disorder-broadening is unknown. Also, our theory includes only the physics of Anderson localization, which would surely be modified by the presence of interaction in the problem. Predicting $\nu_c$ accurately in realistic samples including disorder and interaction microscopically (and allowing the possibility for all possible quantum phases—FQHL, WC, AS, \dots) is completely out of scope for any current theoretical tools. 
We believe that the theories based just on energetics without including disorder miss an essential ingredient of the experiments in real samples. We do emphasize that our purely disorder-based simplified theory predicts the correct experimental trends in $\nu_c$, namely, the decrease in $\nu_c$ with increasing mobility.

We emphasize that both interaction and disorder are important in the LLL physics, as is well-known and also is obvious from the appearance of both FQHE and WC in the results of Ref.~\cite{Yazdani2024}.  
We are by no means saying that interaction can be ignored. 
We are, however, claiming that the uncritical assertion of associating the strongly insulating state at low LLL filling, often observed in transport measurements, automatically with a WC formation may have to be re-evaluated in view of the results of Ref.~\cite{Yazdani2024}.  
We believe that the formation of the strongly insulating state has a lot to do with the amount of disorder in the system associated with the inevitable existence of random charged impurities even in the best 2D samples.  
In particular, Imry-Ma arguments~\cite{Imry1975}  rule out the existence of long range order in the presence of quenched random impurities which act as a random field.  
Therefore, the WC formation in the strict sense is not allowed.  
In fact, there is no spontaneous symmetry breaking here at all since the starting Hamiltonian with random charged impurities does not have any translational order already-- translational symmetry is broken explicitly by the random field associated with disorder and not spontaneously by interaction.  The only question is the extent of the short range order associated with the local (not global) breaking of the rotational symmetry apparent in Fig. 2 of Ref.~\onlinecite{Yazdani2024} for $\nu>0.093$.  Since eventually the local hexagonal pattern disappears with the structure factor manifesting only a ring for $\nu=\nu_c\sim 0.093$ in Ref.~\onlinecite{Yazdani2024}, it is reasonable to assume that the system is an amorphous AS for $\nu<\nu_c$ instead of being any type of Wigner solid.  Of course, a highly fragmented (through strong pinning by many random impurities) strongly pinned Wigner crystal with no spatial ordering is the same as the amorphous AS, and it may then simply be semantics to call it AS or strongly pinned WS.  All we are saying is that it may be misleading to emphasize only the WC aspect of the insulating phase over the AS aspect, particularly since the best theoretical estimates seem to indicate that in a pristine system, the FQHL may remain the ground state for arbitrarily low LLL filling in the absence of any disorder~\cite{Peterson2003,Zuo2020}.


\begin{acknowledgments}
Authors are particularly grateful to Ali Yazdani and Yen-Chen Tsui for nuanced discussions on Ref.~\onlinecite{Yazdani2024}. Authors are also grateful to Herb Fertig, Jainendra Jain, Jay Deep Sau, Mansour Shayegan, and Dan Tsui for extended discussions on the physics of Wigner crystal over many years. This work is supported by the Laboratory for Physical Sciences through its support of the Condensed Matter Theory Center at the University of Maryland. 
\end{acknowledgments}

\appendix

\section{Derivation of Density of States in the Self-Consistent Born Approximation}

We consider a two-dimensional electron gas (2DEG) subjected to a strong perpendicular magnetic field $\mathbf{B} = B\hat{z}$ and a random disorder potential. The Hamiltonian is given by $H = H_0 + V(\mathbf{r})$, where $H_0 = (\mathbf{p} + e\mathbf{A})^2/2m$ is the clean kinetic Hamiltonian with Landau level eigenstates $\ket{n, k}$ and energies $E_N = (N + 1/2)\hbar\omega_c$. The potential $V(\mathbf{r}) = \sum_j u_0 \delta(\mathbf{r} - \mathbf{R}_j)$ represents short-range impurities located at random positions $\mathbf{R}_j$ with concentration $n_{i}$.

The disorder-averaged retarded Green's function is diagonal in the Landau level index $N$ due to the restoration of translational invariance after averaging, given by $G_N(E) = [G_N^{(0)}(E)^{-1} - \Sigma_N(E)]^{-1} = [E - E_N - \Sigma_N(E) + i0^+]^{-1}$, where $G_N^{(0)}(E) = [E - E_N + i0^+]^{-1}$. Using SCBA, we sum over all non-crossing diagrams as shown in Fig.~\ref{fig:LL_SCBA} (b), leading to a self-consistency equation where the internal electron propagator is the full Green's function $G$. For short-range scatterers, the scattering potential is constant in momentum space ($|u(\mathbf{q})|^2 = u_0^2$), simplifying the self-energy to $\Sigma(E) = n_{i} u_0^2 \Tr [G(E)]$.

Assuming the high-field limit where Landau level mixing is negligible ($\hbar\omega_c \gg \Gamma$), we restrict the trace to a single Landau level $n$. The trace over intra-level quantum numbers yields the degeneracy $N_L = (1/2\pi l_B^2)$, leading to the self-consistency condition:
\begin{equation} \label{eq:scba_appendix}
    \Sigma_N(E) = \frac{n_{i} u_0^2}{2\pi l_B^2} G_N(E).
\end{equation}
Defining the disorder strength parameter $\Gamma$ such that
\begin{align}\label{eq:gamma_appendix}
    \frac{\Gamma^2}{4} \equiv \frac{n_{i} u_0^2}{2\pi l_B^2}
\end{align}
and substituting Eq. (\ref{eq:scba_appendix}) into the Green's function definition, we obtain:
\begin{equation}
    \Sigma_N(E) = \frac{\Gamma^2}{4} \frac{1}{E - E_N - \Sigma_N(E)}.
\end{equation}
Letting $\Delta E = E - E_N$, this rearranges to the quadratic equation $\Sigma_N^2 - \Delta E \Sigma_N + \Gamma^2/4 = 0$. Solving for $\Sigma_N$ and selecting the branch with a negative imaginary part within the band ($|\Delta E| < \Gamma_0$) to satisfy causality gives:
\begin{equation}
    \Sigma_N(E) = \frac{\Delta E}{2} - i \frac{\sqrt{\Gamma^2 - (E - E_N)^2}}{2}.
\end{equation}

The DOS per unit area of the $N$-th LL is related to the imaginary part of the Green's function by $D_N(E) = -(1/\pi) (1/2\pi l_B^2) \Im G_N(E)$. Using the relation $G_N(E) = (4/\Gamma^2) \Sigma_n(E)$, we find $\Im G_N(E) = -(2/\Gamma^2) \sqrt{\Gamma^2 - (E - E_n)^2}$. Substituting this back into the expression for $D_N(E)$ yields the semi-elliptic density of states:
\begin{equation}
    D_N(E) = \frac{1}{2\pi l_B^2} \frac{2}{\pi \Gamma} \sqrt{1 - \left( \frac{E - E_N}{\Gamma} \right)^2}.
\end{equation}
This function describes a semi-ellipse centered at $E_N$ with a broadening half width of $\Gamma$, which leads to Eq.~\eqref{eq:dos_scba} in the main text.

For short-range impurity scattering modeled by the potential, the zero-field scattering rate $\tau^{-1}$ is determined by Fermi's Golden Rule using the constant density of states $D_0 = m/(2\pi\hbar^2)$. This yields the expression $\hbar/\tau = n_{i} u_0^2 m/\hbar^2$, where $n_{i}$ is the impurity concentration. This zero-field scattering parameter provides a direct physical interpretation for the Landau level broadening width $\Gamma$ derived in SCBA. By substituting the expression for $\hbar/\tau$ into Eq.~\eqref{eq:gamma_appendix}, we obtain the relation $\Gamma = \sqrt{\frac{2}{\pi} \hbar \omega_c \frac{\hbar}{\tau}}$ (c.f. Eq.~\eqref{eq:gamma_scba} in the main text).


%

\end{document}